\documentclass[aip,graphicx,reprint]{revtex4-1}

\usepackage{amsmath,amssymb}
\usepackage[utf8]{inputenc}
\usepackage{booktabs}
\usepackage{xcolor}
\usepackage{hyperref}
\usepackage{graphicx}

\begin{document}

\title{Physics-Constrained Learning of Dose-Dependent Spectral Degradation\\
in Metal--Organic Frameworks from In Situ Low-Loss EELS}

\author{Gabriel T. dos Santos}
\affiliation{Department of Materials Science and Engineering, Northwestern University, Evanston, IL 60208, USA}

\author{Roberto dos Reis}
\email{roberto.reis@northwestern.edu}
\affiliation{Department of Materials Science and Engineering, Northwestern University, Evanston, IL 60208, USA}
\affiliation{The NU\textit{ANCE} Center, Northwestern University, Evanston, IL 60208, USA}
\affiliation{International Institute of Nanotechnology, Northwestern University, Evanston, Illinois 60208, USA}

\author{Vinayak P. Dravid}
\email{v-dravid@northwestern.edu}
\affiliation{Department of Materials Science and Engineering, Northwestern University, Evanston, IL 60208, USA}
\affiliation{The NUANCE Center, Northwestern University, Evanston, IL 60208, USA}
\affiliation{International Institute of Nanotechnology, Northwestern University, Evanston, Illinois 60208, USA}

\date{\today}

\begin{abstract}
Electron-beam irradiation limits atomic-resolution characterization of
beam-sensitive hybrid materials, yet quantitative models that connect
in situ spectroscopy to dose-dependent degradation remain scarce. Here
we use a physics-informed neural network (PINN) to model beam-induced spectral
evolution in MIL-101(Fe) from an in situ low-loss electron energy-loss
spectroscopy (EELS) dose series. Each spectrum is reduced to fixed-window
low-loss descriptors,
$\tilde n_{\mathrm{eff},j}(\Phi)=\int_{\mathcal{W}_j}S(E,\Phi)\,dE$,
evaluated over nominal $\pi$--$\pi^{*}$, C--C, C--O, and M--O windows. These
descriptors are relative window-integrated low-loss spectral areas, not absolute
f-sum-rule effective electron numbers. For each spectral channel, a latent
integrity variable $C_i(\Phi)$ obeys the same uncoupled power-law degradation
equation in normalized dose space, $dC_i/d\phi=-k_i C_i^{p_i}$, regularized by
monotonicity, boundedness, and a single hierarchy prior
$k_{\mathrm{C\text{-}O}}\geq k_{\mathrm{C\text{-}C}}$. Applied to nine dose
frames spanning 152--1368~e$^-$/\AA$^2$, the ensemble PINN identifies C--O and
C--C as the most strongly dose-sensitive linker-associated channels, with
half-integrity thresholds of approximately $1.0\times10^3$~e$^-$/\AA$^2$. The
1--3~eV $\pi$--$\pi^{*}$-labelled window increases with dose and is therefore
interpreted as a mixed low-energy response, likely involving
oscillator-strength redistribution rather than direct monotonic loss of a
single bond population. The framework provides a dose-dependent,
spectroscopy-constrained description of MOF degradation while explicitly stopping short of chemical-mechanism assignment: fixed-window low-loss EELS descriptors are phenomenological and require additional chemical-state validation for assigned to specific bond populations. 
\end{abstract}


\maketitle

\section{Introduction}
\label{sec:intro}

Metal--organic frameworks (MOFs) are hybrid crystalline materials whose
tunable porosity and chemical functionality make them attractive for catalysis,
gas storage, and separations.\cite{Furukawa2013,Zhou2012} Their
characterization by transmission electron microscopy (TEM), scanning
transmission electron microscopy (STEM), electron diffraction, and ptychographic
imaging is difficult because many MOFs lose crystallinity or undergo chemical
modification at low electron
dose.\cite{Chen2020,Zhan2024,Liu2019,Xu2025,Li2025,Ghosh2019,Gnanasekaran2024,Banihashemi2020,Tien2024}
For operando
measurements this creates a direct trade-off: the electron beam provides the
structural or spectroscopic signal, but the same beam can drive radiolysis,
charging, knock-on displacement, and framework reconstruction. Existing studies
have established dose thresholds and macroscopic damage signatures through
diffraction, imaging, 4D-STEM, and EELS, but they do not by themselves provide
a compact kinetic model for how different spectral components evolve with
cumulative dose.

Physics-informed neural networks (PINNs) offer a useful route for this problem
because ordinary differential equations and inequality constraints can be
embedded directly into the loss function while fitting sparse, noisy
experimental data.\cite{Raissi2019,Karniadakis2021} Here we use low-loss EELS
as the input signal because the low-loss region is dose efficient and sensitive
to changes in the valence-electron response.\cite{Egerton2004,Egerton2011,Egerton2013,Egerton2019} We
do not claim that the fixed energy windows used here uniquely isolate
individual chemical bonds. Instead, we use nominally labelled low-loss windows
($\pi$--$\pi^{*}$, C--C, C--O, and M--O) as phenomenological spectral
descriptors and ask whether a constrained dose-space PINN can recover stable,
interpretable trends from the measured dose series. Our main contribution is a
channel-wise degradation model for MIL-101(Fe) that identifies C--O and C--C as
the most dose-sensitive linker-associated responses, while explicitly treating
the low-energy $\pi$--$\pi^{*}$-labelled window as a mixed response that
requires separate physical caution. This contribution is intentionally methodological rather than mechanistic: the framework is designed to extract physically-constrained, generalizable kinetic descriptors from sparse dose series, not to resolve the underlying degradation chemistry in full.

\section{Experimental Data: In Situ Low-Loss EELS}
\label{sec:experimental}

\subsection{Acquisition}
In situ low-loss EELS spectra were acquired as a nine-frame cumulative dose
series spanning 152--1368~e$^-$/\AA$^2$ at room temperature using a Gatan
Continuum GIF with K3 direct electron detector coupled to a JEOL GRAND ARM
300F, operated at 300~kV. Dual EELS collection was performed with an energy
dispersion of 0.18~eV/ch; each frame was collected sequentially every 44~s over
a total acquisition time of 6:42~min (402~s). The full width at half-maximum
(FWHM) of the zero-loss peak (ZLP) was 0.6~eV.

\subsection{Pre-processing and low-loss window integration}
\label{sec:preprocess}
Low-loss EELS spectra were processed using Gatan Microscopy Suite (GMS) and
custom scripts. The processing workflow included energy calibration with the
zero-loss peak, removal of X-ray spikes, background subtraction with a
power-law model, Fourier-log deconvolution to reduce plural scattering, and
normalization to the integrated zero-loss intensity. The resulting processed
low-loss spectra, denoted $S(E,\Phi)$, were used directly for
fixed-window integration. No Kramers--Kronig inversion or f-sum-rule
normalization was applied in the PINN input pipeline.

\subsection{Low-loss descriptors and energy windows}
\label{sec:observables}
Let $y_j(\Phi)$ denote the experimentally measured descriptor for energy
window $j$ at accumulated dose $\Phi$. We define
\begin{equation}
y_j(\Phi) \equiv \tilde n_{\mathrm{eff},j}(\Phi)
= \int_{E\in\mathcal{W}_j} S(E,\Phi)\,dE ,
\label{eq:window_integral}
\end{equation}
where $S(E,\Phi)$ is the processed low-loss EELS intensity and the integral is evaluated
by trapezoidal quadrature. The tilde indicates that this is a relative
window-integrated low-loss spectral area. It is not an absolute effective electron number
obtained from Kramers--Kronig analysis or f-sum-rule normalization. The four
windows used in the analysis are $\pi$--$\pi^{*}$ labelled (1--3~eV), C--C
(4--7~eV), C--O (10--15~eV), and M--O (20--25~eV). These labels are shorthand
for fixed spectral windows and should not be read as unique bond assignments.

\section{Dose-Dependent Degradation Model}
\label{sec:model}

\subsection{Latent integrity variables}
We define latent integrity variables $C_i(\phi)$ representing the relative
integrity of channel $i \in \{\pi\text{-}\pi^{*},\;\mathrm{C\text{-}C},\;\mathrm{C\text{-}O},\;
\mathrm{M\text{-}O}\}$:
\begin{equation}
0 \le C_i(\phi) \le 1,\qquad C_i(0)=1.
\label{eq:integrity}
\end{equation}

\subsection{Dose-driven kinetics}
Each channel is described by the same uncoupled power-law ordinary
differential equation in normalized dose space,
\begin{equation}
\frac{dC_i}{d\phi} = -k_i\, C_i^{\,p_i},
\label{eq:ode}
\end{equation}
where $\phi=(\Phi-\Phi_0)/(\Phi_{\max}-\Phi_0)\in[0,1]$, $k_i>0$ is the
apparent channel degradation rate, and $p_i\geq1$ is the kinetic exponent. The
same functional form is used for all four channels. No inter-channel coupling,
recombination term, or graphitization source term is included in the current
model.

\subsection{Hierarchy prior}
The only hierarchy prior imposed during training is the soft pairwise
constraint
\begin{equation}
k_{\mathrm{C\text{-}O}} \ge k_{\mathrm{C\text{-}C}}.
\label{eq:hierarchy}
\end{equation}
No hierarchy is imposed on the $\pi$--$\pi^{*}$-labelled or M--O channels.

\begin{figure*}[t]
\centering
\includegraphics[width=0.85\textwidth]{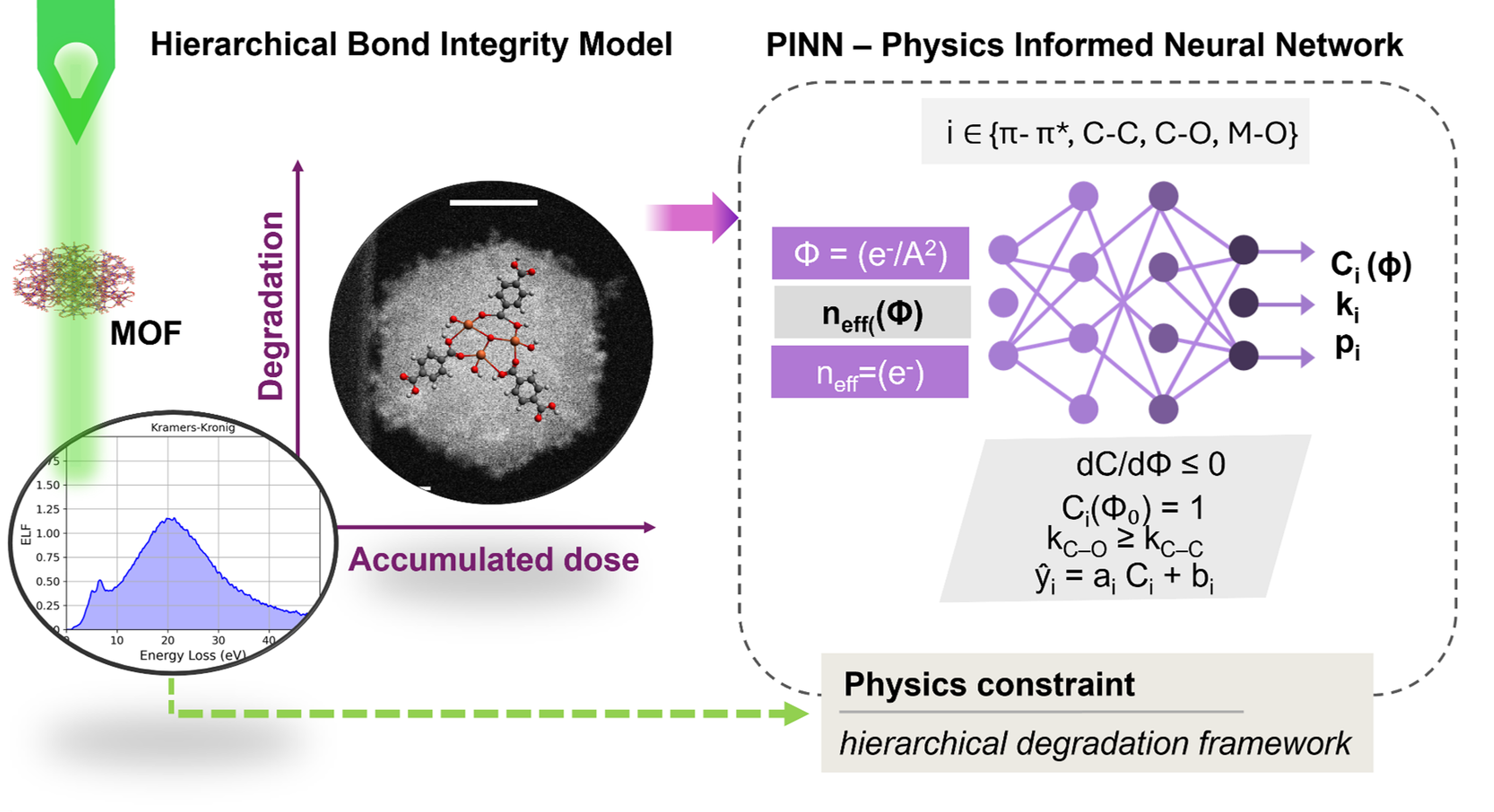}
\caption{Schematic illustration of the PINN-based framework for modeling beam-induced spectral evolution in MOFs. In situ low-loss EELS data are processed into fixed-window low-loss descriptors, which are used to train a physics-informed neural network. The model infers latent channel trajectories $C_i(\Phi)$ and apparent kinetic parameters ($k_i$, $p_i$) under monotonicity, boundedness, and hierarchy regularization. Scale bar 200~nm.}
\label{fig:overview}
\end{figure*}

\section{Physics-Informed Neural Network}
\label{sec:pinn}

\subsection{Network architecture}
The PINN takes normalized dose $\phi = (\Phi -\Phi_0)/(\Phi_{\max} - \Phi_0) \in [0,1]$ as input and predicts latent integrity states:
\begin{equation}
\widehat{\mathbf{C}}(\phi) = \mathrm{NN}_\theta(\phi).
\end{equation}
The network employs random Fourier feature (RFF) encoding \cite{Tancik2020} to overcome spectral bias, followed by a modified multilayer perceptron (MLP) with U-V encoder skip connections \cite{Wang2021} across five hidden layers (64~neurons each; 29,396 trainable parameters). A reparameterized output head enforces the boundary condition $C_i(\Phi_0)=1$ by construction. The overall framework is schematically illustrated in Fig.~1.

\subsection{Link functions}
An affine readout connects each latent channel trajectory to its measured
low-loss descriptor,
\begin{equation}
\widehat{y}_j(\Phi) = a_j\,\widehat{C}_j(\Phi) + b_j,
\label{eq:link}
\end{equation}
The coefficient $a_j$ is not sign constrained in the notebook pipeline used to
generate the manuscript figures. A positive $a_j$ maps decreasing latent
integrity to decreasing window intensity. A negative $a_j$ allows a
monotonically decreasing latent state to represent a window whose measured
intensity increases with dose, as occurs for the 1--3~eV
$\pi$--$\pi^{*}$-labelled channel. In that case $C_j(\Phi)$ should be
interpreted as a latent monotonic coordinate, not as a direct electron count for
a single bond population.

\subsection{Loss function}
The total loss combines data fidelity, physics residuals, and constraints:
\begin{equation}
\mathcal{L} = \lambda_{\mathrm{data}}\,\mathcal{L}_{\mathrm{data}}
            + \lambda_{\mathrm{phys}}\,\mathcal{L}_{\mathrm{phys}}
            + \lambda_{\mathrm{con}}\,\mathcal{L}_{\mathrm{con}}.
\label{eq:loss_total}
\end{equation}

\paragraph{Data fidelity.}
The data term is evaluated as a log-cosh loss between z-scored experimental
descriptors and model predictions,
\begin{equation}
\mathcal{L}_{\mathrm{data}} =
\frac{1}{N}\sum_{t=1}^{N}\sum_{j}
\log\!\left[\cosh\!\left(\widehat{y}_j(\Phi_t)-y_j(\Phi_t)\right)\right],
\end{equation}
with a threefold weight applied to the $\pi$--$\pi^{*}$-labelled channel to
prevent the low-magnitude low-energy window from being ignored during
optimization.

\paragraph{Physics residuals.}
Using automatic differentiation:
\begin{equation}
\mathcal{L}_{\mathrm{phys}} =
\frac{1}{N}\sum_{t=1}^{N}\sum_{i}
\left(
\frac{d\widehat{C}_i}{d\phi}\bigg|_{\phi_t}
+ k_i\,\widehat{C}_i(\Phi_t)^{p_i}
\right)^2.
\end{equation}

\paragraph{Constraints.}
\begin{align}
\mathcal{L}_{\mathrm{con}} &=
\sum_{t,i}\!\left[
  \mathrm{ReLU}(-\widehat{C}_i)^2
  + \mathrm{ReLU}(\widehat{C}_i-1)^2
\right]
\nonumber\\
&\quad + \sum_{t,i}\mathrm{ReLU}\!\!\left(
  \frac{d\widehat{C}_i}{d\phi}\bigg|_{\phi_t}
\right)^{\!2}
+ \mathcal{L}_{\mathrm{hier}},
\label{eq:loss_con}
\end{align}
where $\mathcal{L}_{\mathrm{hier}} = \mathrm{ReLU}(k_{\mathrm{C\text{-}C}} -
k_{\mathrm{C\text{-}O}})^2$ penalizes hierarchy violations.

\subsection{Training protocol}
All manuscript figures and kinetic parameters were generated from the notebook
pipeline. The network uses a 32-frequency random Fourier feature embedding of
normalized dose,\cite{Tancik2020} followed by a modified MLP with U--V encoder
skip connections.\cite{Wang2021} The initial condition $C_i(0)=1$ is enforced
by the hard parameterization $C_i(\phi)=1-\phi\sigma[g_i(\phi)]$. Five ensemble
members were trained on CPU with seeds 42, 59, 76, 93, and 110. Each member was
optimized for 10,000 Adam epochs followed by 2,000 L-BFGS iterations. The
physics residual was evaluated on a dense collocation grid, and causal
weighting was applied uniformly to the channel-summed residual during the Adam
phase.\cite{Wang2022causal,Wang2022ntk} The auxiliary script
\texttt{run\_pinn\_lowloss.py} is a simplified
single-seed reference implementation and was not used to generate the reported
ensemble figures.

\section{Results}
\label{sec:results}

\subsection{PINN fits and inferred integrity trajectories}

Figure~2 summarizes the experimental low-loss EELS dose series. Nine frames
were acquired over 402~s, spanning nominal cumulative doses of
152--1368~e$^-$/\AA$^2$. The processed low-loss spectra were integrated over four
fixed energy windows to generate the channel descriptors used for PINN
training. The C--C, C--O, and M--O windows show modest monotonic or
near-monotonic evolution over the measured dose range. In contrast, the
1--3~eV $\pi$--$\pi^{*}$-labelled window increases with dose. This opposite
trend is important: it means that this low-energy descriptor cannot be
interpreted as direct monotonic loss of a native $\pi$--$\pi^{*}$ bond
population.

\begin{figure}[t]
\centering
\includegraphics[width=0.95\linewidth]{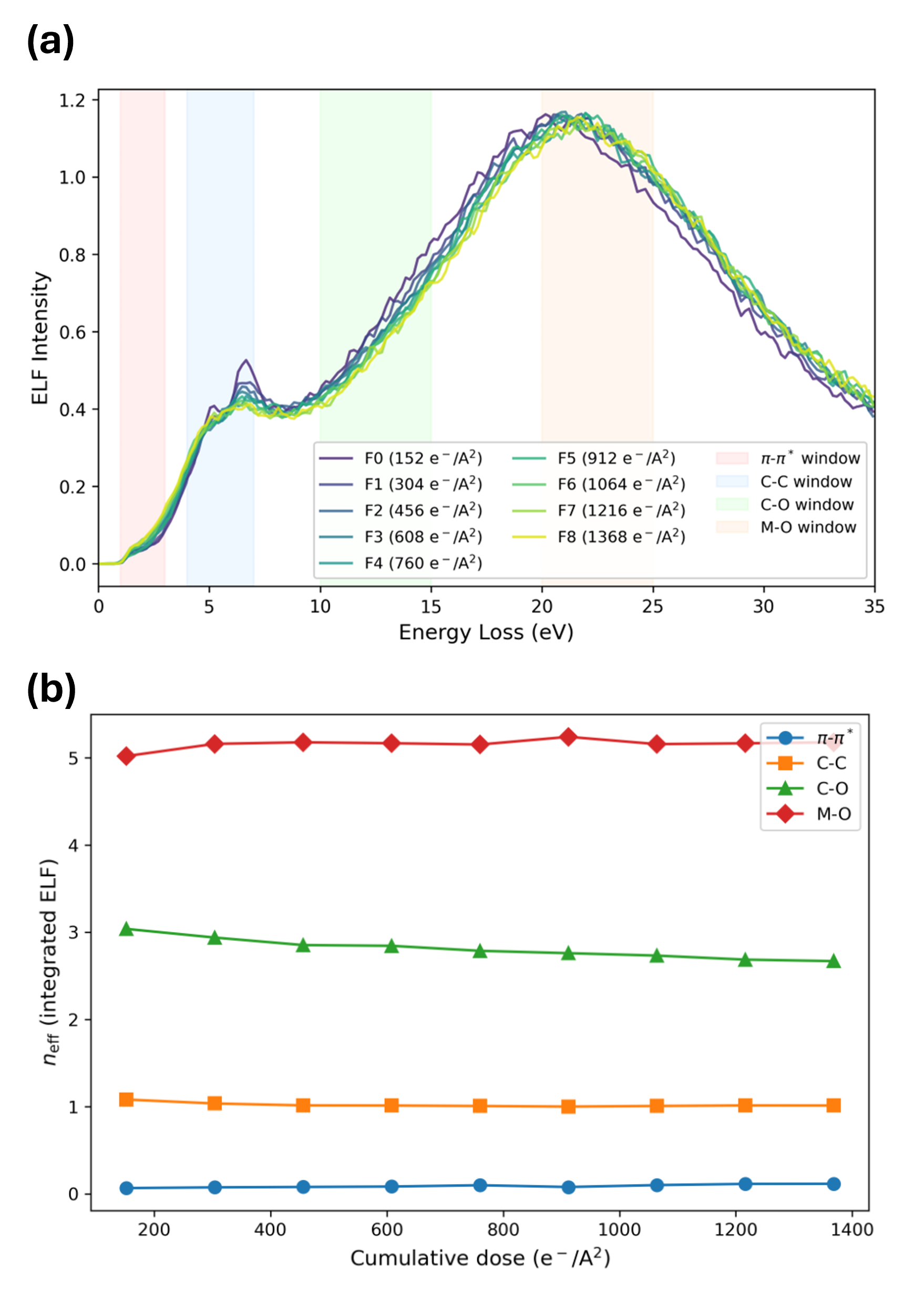}
\caption{(a) In situ low-loss EELS spectra acquired over nine cumulative dose frames (152--1368~e$^-$/\AA$^2$). The shaded regions indicate the fixed integration windows used for the nominal $\pi$--$\pi^{*}$, C--C, C--O, and M--O channels. The strongest spectral evolution occurs in the surface-plasmon region ($\sim$15--25~eV), while the low-energy region shows a smaller but non-negligible dose response. (b) Evolution of the window-integrated low-loss descriptor $\tilde n_{\mathrm{eff}}$ as a function of cumulative dose for each channel. The descriptors are relative window areas, not absolute f-sum-rule effective electron numbers.
}
\label{fig:pinn_results}
\end{figure}

The PINN framework described in Section~IV was fit to the four window
descriptors as a function of cumulative dose. The resulting latent trajectories
$C_i(\Phi)$ are presented in Fig.~3. The C--O and C--C channels show the
largest loss of latent integrity and reach the half-integrity threshold within
the measured dose range. The M--O channel remains weakly varying. The
$\pi$--$\pi^{*}$-labelled channel decreases slowly in latent space even though
its measured window intensity increases; this behavior is enabled by a negative
affine readout coefficient and should be interpreted as a mixed low-energy
response rather than a direct degradation rate for a single $\pi$--$\pi^{*}$
bond population.

\begin{figure}[t]
\centering
\includegraphics[width=0.95\linewidth]{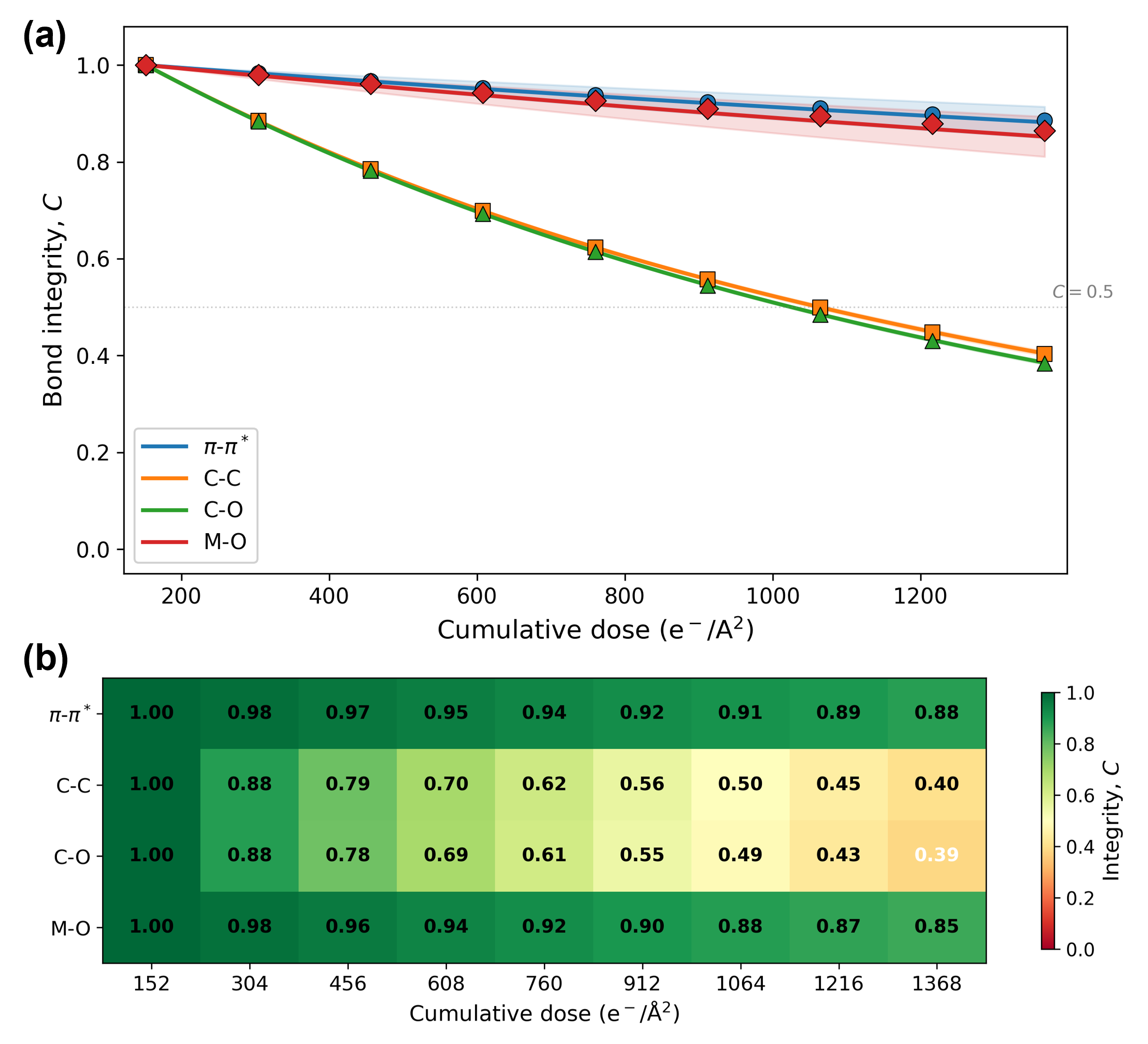}
\caption{(a) PINN-inferred latent channel-integrity trajectories $C_i(\Phi)$ as a function of cumulative electron dose for each spectral channel ($\pi$--$\pi^{*}$ labelled, C--C, C--O, and M--O). Solid lines represent the ensemble mean ($N=5$), with shaded regions indicating $\pm 1\sigma$ uncertainty. The C--O and C--C channels exhibit the largest loss of latent integrity. The $\pi$--$\pi^{*}$-labelled trajectory should be read as a mixed low-energy response because the measured 1--3~eV descriptor increases with dose. (b) Heatmap representation of latent channel-integrity values $C_i$ across the measured dose range (152--1368~e$^-$/\AA$^2$).}
\label{fig:validation}
\end{figure}

Figure~4 summarizes the apparent kinetic parameters inferred from the
five-member ensemble. The C--O and C--C channels form the best-constrained
high-sensitivity group, with
$k_{\mathrm{C\text{-}O}}=0.994\pm0.003$ and
$k_{\mathrm{C\text{-}C}}=0.990\pm0.008$ in normalized-dose units. The
difference between these two rates is small, so the result should be
interpreted as support for a linker-associated high-sensitivity regime, not as
a precise quantitative separation between C--O and C--C scission rates. The
$\pi$--$\pi^{*}$-labelled and M--O channels have lower apparent rates and
larger relative seed-to-seed spread, reflecting weaker identifiability in those
windows.

\subsection{Apparent degradation rates and half-integrity dose} 

Table~\ref{tab:params} summarizes the apparent kinetic parameters for
MIL-101(Fe). We define $\Phi_{1/2,i}$ as the dose at which the latent
coordinate reaches $C_i=0.5$. This threshold is a conventional half-integrity
reference, not a uniquely defined physical phase boundary. For the two
linker-associated channels, the ensemble-mean thresholds are
$\Phi_{1/2,\mathrm{C\text{-}O}}\approx1027$~e$^-$/\AA$^2$ and
$\Phi_{1/2,\mathrm{C\text{-}C}}\approx1063$~e$^-$/\AA$^2$. The
$\pi$--$\pi^{*}$-labelled and M--O channels do not reach $C=0.5$ within the
measured dose range. We therefore report the practical linker-network dose
scale as approximately $1.0\times10^3$~e$^-$/\AA$^2$, rather than a single
sharply defined value.

\begin{figure}[t]
\centering
\includegraphics[width=\linewidth]{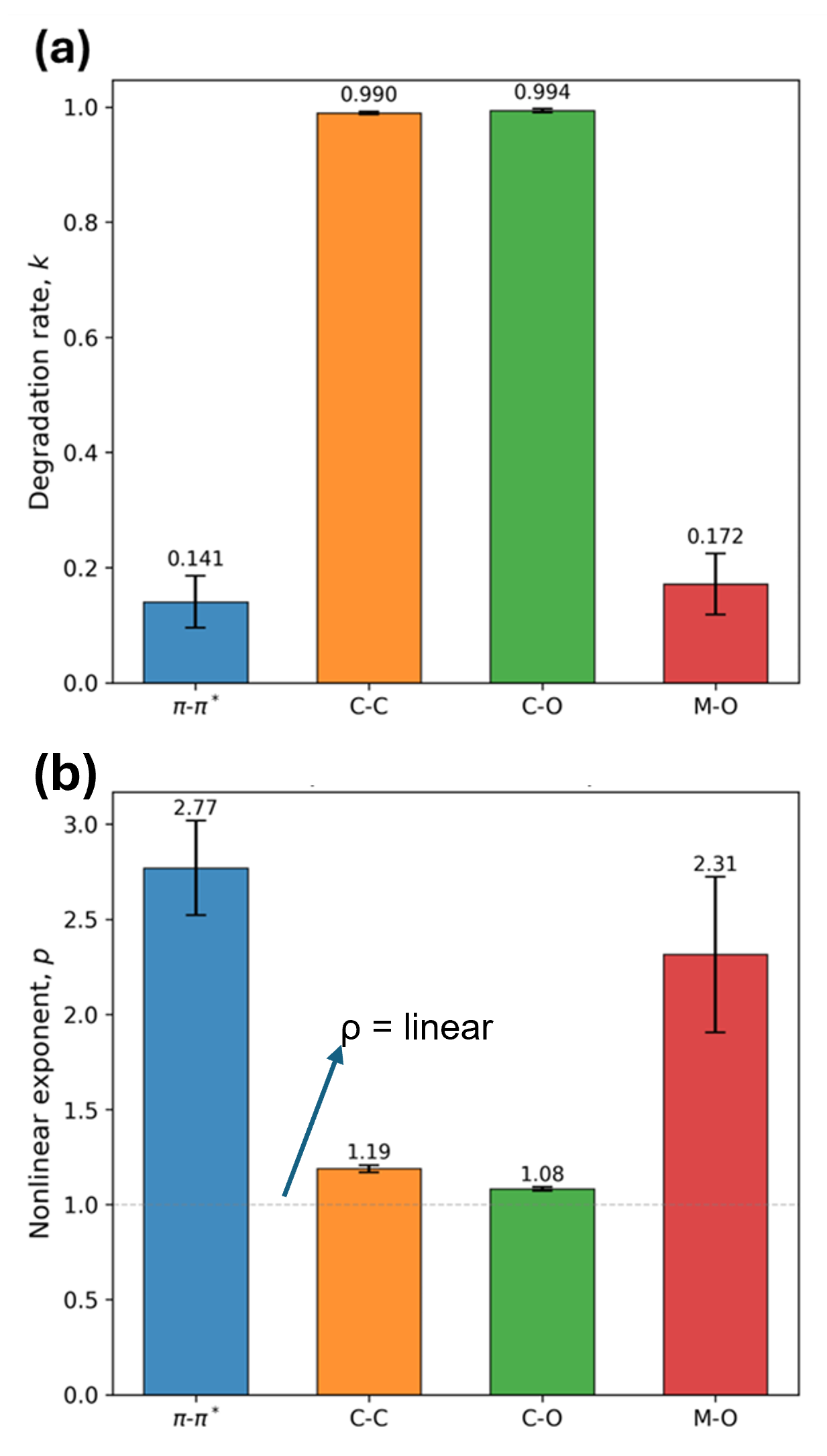}
\caption{(a) Apparent channel degradation rate constants $k_i$ inferred by the PINN ensemble ($N=5$), with error bars representing $\pm 1\sigma$. The C--O and C--C channels define the best-constrained high-sensitivity group. (b) Corresponding nonlinear kinetic exponents $p_i$ for each spectral channel. The dashed line indicates $p = 1$ (linear behavior). The $\pi$--$\pi^{*}$-labelled and M--O channels have larger relative uncertainty and should be interpreted more cautiously.}
\label{fig:ablation}
\end{figure}

\begin{table}[b]
\caption{PINN-inferred apparent kinetic parameters for MIL-101(Fe). Values are
ensemble mean $\pm 1\sigma$ ($N=5$). All channel labels denote fixed spectral windows; the $\pi$--$\pi^{*}$-labelled channel is interpreted as a mixed spectral response.}
\label{tab:params}
\begin{ruledtabular}
\begin{tabular}{lccc}
Channel & $k_i$ & $p_i$ & $\Phi_{1/2}$ (e$^-$/\AA$^2$) \\
\hline
$\pi$--$\pi^{*}$            & $0{.}141 \pm 0{.}020$ & $2{.}77 \pm 0{.}12$ & $>1368$ \\
C--C                       & $0{.}990 \pm 0{.}008$ & $1{.}19 \pm 0{.}05$ & $\sim1063$ \\
C--O                       & $0{.}994 \pm 0{.}003$ & $1{.}08 \pm 0{.}04$ & $\sim1027$ \\
M--O                       & $0{.}172 \pm 0{.}036$ & $2{.}31 \pm 0{.}35$ & $>1368$ \\
\end{tabular}
\end{ruledtabular}
\end{table}

\section{Discussion}
\label{sec:discussion}

\subsection{Interpretability and physical consistency}
\label{sec:interp}

The main physically interpretable result is the separation between the
linker-associated C--O/C--C windows and the more weakly varying
$\pi$--$\pi^{*}$-labelled and M--O windows. The model imposes only one
hierarchy prior, $k_{\mathrm{C\text{-}O}}\geq k_{\mathrm{C\text{-}C}}$, so
the broader separation between linker-associated and weakly varying channels is
not directly hard-coded. At the same time, the C--O and C--C rate constants are
very close in the ensemble fit. We therefore interpret them as a coupled
linker-degradation regime rather than as a robust measurement of a large
kinetic gap between carboxylate and aromatic C--C contributions.

The $\pi$--$\pi^{*}$-labelled channel requires a different level of caution.
The measured 1--3~eV window increases with dose, whereas the latent coordinate
is constrained to be monotonic. The fit is therefore carried by the affine
readout, not by a direct monotonic decrease of the measured low-energy
intensity. A plausible interpretation is that electron irradiation redistributes
oscillator strength into the low-energy region as the organic linker is
damaged, possibly through carbonaceous or more conjugated fragments. EELS
studies of carbonaceous materials show that carbonization and graphitization can
be accompanied by changes in low-loss and near-edge spectral
features,\cite{Daniels2007} and MOF beam damage is known to involve coupled
radiolytic and structural processes rather than a single isolated bond-breaking
coordinate.\cite{Ghosh2019,Gnanasekaran2024,Banerjee2024} This interpretation is further supported by the core-loss EELS comparison in the Supplementary Material (Fig. S3), which show a loss of definition of the carboxylate-associated $pi^*_2$ feature relative to the aromatic-ring $pi^*_1$ feature with increasing dose. An additional, non-exclusive electronic origin should also be considered. MIL-101(Fe) has an optical band gap of approximately 2.5~eV assigned to ligand-to-metal charge transfer (LMCT) transitions between the carboxylate linker and the Fe(III) oxo-clusters~\cite{li2025bandgap} This interpretation remains phenomenological. The present model does not fit a graphitization rate, does not include a source term, and does not prove that recombination or graphitization is localized exclusively in the 1--3~eV window. The uncoupled, per-channel formulation adopted here (Eq.~3) is a simplifying assumption, and two aspects of the present results indicate its limits. The near-degenerate rate constants obtained for the C--O and C--C channels ($k_{C-O} = 0.994 \pm 0.003$ vs. $k_{C-C} = 0.990 \pm 0.008$) suggest shared upstream degradation kinetics rather than fully independent evolution. More directly, the $\pi$--$\pi^*$-labelled channel is poorly described by a monotonic decay law of any kind: it requires a sign-inverted affine readout to fit a non-monotonic observable, and the unconstrained neural-network baseline achieves only $R^2 = 0.006$ for this channel (Fig.~S4a), indicating that its dose dependence is not well captured by a simple monotonic functional form, constrained or otherwise. The PINN was compared to three standard nonlinear baselines (single exponential, bi-exponential, and stretched exponential) for each channel using small-sample-corrected model selection criteria (Table~S1, Supplementary Material); the single exponential is favored across all four channels, confirming that the observed dose-dependent trends are well described by conventional models within the present sampling, and that the value of the physics-informed formulation lies specifically in constrained extrapolation (Fig.~S4a) rather than in-domain fit quality. We identify the development of a coupled-channel formulation, together with a dedicated source-term model for the $\pi$--$\pi^*$ channel (potentially incorporating dose-induced oscillator-strength redistribution) and disentangling carbonization from LMCT/oxidation-state contributions to the 1–3 eV window, as priority directions for future work.

\subsection{What low-loss EELS can and cannot capture}
\label{sec:limitations}

Low-loss EELS offers a dose-efficient way to monitor changes in the
valence-electron response during beam exposure, but the fixed-window analysis
used here has clear limits. First, the descriptors are spatially averaged over
the illuminated region and cannot resolve surface-initiated or intraparticle
degradation fronts. Second, the windows are not element-specific. Changes in Fe
oxidation state, oxygen coordination, or ligand chemistry would be more
directly tested by core-loss EELS, X-ray spectroscopy, Raman spectroscopy,
diffraction, or simulated EEL spectra. A preliminary qualitative comparison of core loss C--K, O--K, and Fe--L$_2,_3$ spectra at the lowest- and highest-dose is provided in the Supplementary Material (Fig S3), consistent with a linker-dominated, metal-node-preserving degradation pathway. A full quantitative core-loss dose series remains a priority for future work. Third, the 1--3~eV $\pi$--$\pi^{*}$-labelled descriptor should be treated as a low-energy mixed-response window rather than a unique aromatic $\pi$--$\pi^{*}$ electron count. Finally, the reported kinetic parameters are effective descriptors under the present 300~kV, room-temperature, nominal-dose conditions; they should not be transferred to other microscope conditions without dose-rate, voltage, thickness, and environmental controls. Quantitative access to hydrogen-related vibrations or light-element bonding dynamics would require dedicated monochromated or vibrational EELS experiments under low-dose conditions.\cite{Krivanek2019,Krivanek2014} The broader radiation-damage limits and mitigation strategies for TEM therefore remain central to applying this workflow beyond the present experiment.\cite{Egerton2004,Egerton2013}

\subsection{Outlook: spatial PINN and multi-condition extension}
\label{sec:outlook}

The framework presented here treats dose as the sole independent variable, yielding spatially averaged kinetics. A natural extension is to promote the integrity variables to spatiotemporal fields, $C_i(\mathbf{r}, \Phi)$, trained on EELS spectrum-image (SI) dose series or 4D-STEM hyperspectral data. Such a \emph{spatial PINN} would enable mapping of degradation fronts within individual particles, connecting mesoscale morphological collapse to bond-level kinetics.
Beyond spatial resolution, the dose-space ODE (Eq.~\ref{eq:ode}) can be extended to incorporate additional experimental variables---temperature, atmosphere, and dose rate---as co-independent inputs, transforming the model into a multi-condition predictor of radiation tolerance. Comparative studies across isoreticular MOF families (e.g., MIL-101(Cr), UiO-66, ZIF-8) would allow direct assessment of how metal-node chemistry and linker topology modulate degradation pathways. Finally, coupling the low-loss PINN with core-loss EELS channels (e.g., Fe~$L_{2,3}$, O~$K$) in a multi-task learning framework would provide simultaneous access to both electronic structure evolution and element-specific oxidation-state kinetics, yielding a more complete picture of radiation damage in hybrid materials.

\section{Conclusions}
\label{sec:conclusions}

We show that fixed-window low-loss EELS descriptors can be combined with a
constrained dose-space PINN to extract effective degradation trends in
MIL-101(Fe). The most robust result is that the C--O and C--C linker-associated
windows form the high-sensitivity regime, reaching $C=0.5$ at approximately
$1.0\times10^3$~e$^-$/\AA$^2$ under the nominal dose calibration used here. The
M--O window remains weakly varying over the measured range. These trends
converge with qualitative observations from the core-loss region. The 1--3~eV
$\pi$--$\pi^{*}$-labelled window behaves differently: its measured intensity
increases with dose, so its inferred latent trajectory should be read as a
phenomenological mixed-response coordinate rather than a direct degradation
rate for a single bond population. This distinction is essential for using the
model correctly. The framework is useful as a spectroscopy-constrained
degradation descriptor, but chemical assignment of the individual windows will
require additional validation by Raman spectroscopy, diffraction, controlled carbonization references, or simulated EEL spectra.

\section*{Supplementary Material}

The supplementary material provides definitions of the 
window-integrated low-loss descriptors, per-channel PINN 
ensemble fits (Fig.~S1), training diagnostics (Fig.~S2), a core-loss EELS data (Fig.~S3), and an ablation study against unconstrained baselines (Fig.~S4).

\begin{acknowledgments}
This work was primarily supported as part of the Hydrogen in Energy and
Information Sciences (HEISs) Center, an Energy Frontier Research Center funded
by the U.S.\ Department of Energy, Office of Science, Basic Energy Sciences
under Award No.\ DE-SC0023450. This work made use of the EPIC facility
(RRID:~SCR\_026361) of Northwestern University's NU\textit{ANCE} Center, which has
received support from the SHyNE Resource (NSF ECCS-2025633), the International
Institute of Nanotechnology (IIN), and Northwestern's MRSEC program
(NSF DMR-2308691). Computational resources were provided by the Quest High
Performance Computing Facility at Northwestern University.
\end{acknowledgments}

\section*{Author Declarations}

\subsection*{Conflict of Interest}
The authors declare no competing interests.

\subsection*{Author Contributions}
\textbf{Gabriel T. dos Santos:} Conceptualization, Methodology, Software,
Formal Analysis, Investigation, Data Curation, Visualization,
Writing -- Original Draft.
\textbf{Roberto dos Reis:} Conceptualization, Methodology, Supervision,
Writing -- Review \& Editing.
\textbf{Vinayak P. Dravid:} Resources, Supervision, Funding Acquisition,
Writing -- Review \& Editing.

\subsection*{Data Availability}
The data that support the findings of this study are available from the
corresponding author upon reasonable request.

\subsection*{Code Availability}
The full analysis framework, including EELS pre-processing scripts,
fixed-window low-loss integration routines, and the physics-informed neural network
implementation for dose-dependent spectral degradation modeling, is available
at \url{https://github.com/gabrielgts007/hierarchical-pinn-elf}. The manuscript
figures and ensemble kinetic table were generated from the notebook workflow.
The auxiliary script \texttt{run\_pinn\_lowloss.py} is a simplified reference
implementation and is not the canonical source of the reported figures.

\bibliographystyle{aipnum4-2}
\bibliography{references}

\end{document}


\fi

\title{Supplementary Material:\\
Physics-Constrained Learning of Dose-Dependent Spectral Degradation
in Metal--Organic Frameworks from In Situ Low-Loss EELS}

\author{Gabriel T. dos Santos}
\affiliation{Department of Materials Science and Engineering,
Northwestern University, Evanston, IL 60208, USA}

\author{Roberto dos Reis}
\email{roberto.reis@northwestern.edu}
\affiliation{Department of Materials Science and Engineering,
Northwestern University, Evanston, IL 60208, USA}
\affiliation{The NUANCE Center, Northwestern University,
Evanston, IL 60208, USA}
\affiliation{International Institute of Nanotechnology,
Northwestern University, Evanston, IL 60208, USA}

\author{Vinayak P. Dravid}
\email{v-dravid@northwestern.edu}
\affiliation{Department of Materials Science and Engineering,
Northwestern University, Evanston, IL 60208, USA}
\affiliation{The NUANCE Center, Northwestern University,
Evanston, IL 60208, USA}
\affiliation{International Institute of Nanotechnology,
Northwestern University, Evanston, IL 60208, USA}

\maketitle
\renewcommand{\thefigure}{S\arabic{figure}}
\setcounter{figure}{0}
\vspace{-1.5em}

\section{Window-integrated low-loss descriptors}
\label{sec:window_descriptors}

The PINN input descriptors were calculated by integrating the processed
low-loss EELS intensity over fixed energy windows.  For each frame and
channel,
\begin{equation}
\tilde n_{\mathrm{eff},j}(\Phi)=
\int_{E\in\mathcal{W}_j}S(E,\Phi)\,dE ,
\label{eq:window_integral_si}
\end{equation}
where $S(E,\Phi)$ is the processed low-loss spectrum and the integral is
evaluated by trapezoidal quadrature.  The notation
$\tilde n_{\mathrm{eff}}$ is used only as a relative window-area
descriptor.  These values are not absolute f-sum-rule effective electron
numbers and were not obtained by Kramers--Kronig inversion.

The four integration windows were fixed before training:
$\pi$--$\pi^{*}$ labelled (1--3~eV), C--C (4--7~eV), C--O (10--15~eV),
and M--O (20--25~eV).  These assignments are phenomenological labels for
spectral windows.  The 1--3~eV window is especially sensitive to
low-energy intensity redistribution and should not be treated as a unique
aromatic $\pi$--$\pi^{*}$ electron count.

\section{Per-Channel PINN Fits}
\label{sec:channel_fits}

\begin{figure*}[p]
\centering
\includegraphics[width=0.85\textwidth]{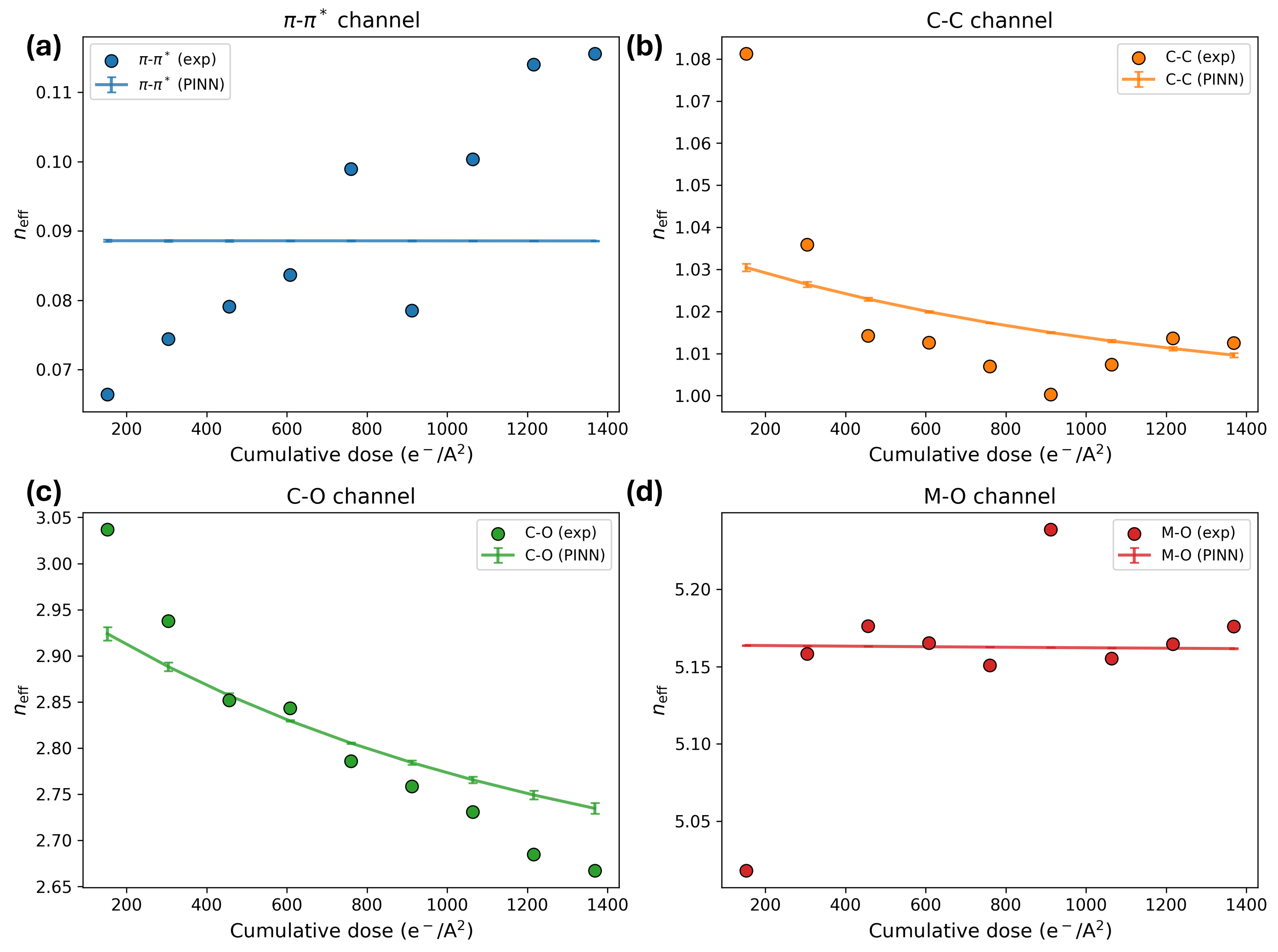}
\caption{Per-channel PINN fits to experimental
$\tilde n_{\mathrm{eff},j}(\Phi)$ descriptors for MIL-101(Fe).
(a)~$\pi$--$\pi^{*}$ (1--3~eV),
(b)~C--C (4--7~eV),
(c)~C--O (10--15~eV),
(d)~M--O (20--25~eV).
Circles: experimental fixed-window low-loss descriptors; solid lines:
ensemble-mean PINN prediction; error bars: $\pm 1\sigma$ over $N = 5$
independent initializations.}
{\small\itshape\raggedright
Alt text: Four scatter plots with fitted curves comparing PINN predictions to experimental low-loss descriptors per channel. The pi-pi* channel rises with dose despite a flat fit; C-C and C-O decline, with C-O most monotonic; M-O stays nearly flat throughout.\par}
\label{fig:S1}
\end{figure*}

Figure~\ref{fig:S1} displays the PINN ensemble predictions
(lines with $\pm 1\sigma$ error bars, $N = 5$) overlaid on the
experimental $\tilde n_{\mathrm{eff},j}(\Phi)$ descriptors (circles) for
each of the four fixed spectral windows.  The C--O channel
[Fig.~\ref{fig:S1}(c)] exhibits the most pronounced monotonic decrease in
the measured descriptor.  The C--C channel [Fig.~\ref{fig:S1}(b)] shows a
similar decreasing trend, albeit with more scatter at intermediate doses.
The $\pi$--$\pi^{*}$-labelled channel [Fig.~\ref{fig:S1}(a)] displays a
non-monotonic measured response, including an increase with dose; the
monotonic latent coordinate should therefore be interpreted as an
effective mixed-response descriptor rather than direct loss of a single
bond population.  The M--O channel [Fig.~\ref{fig:S1}(d)] remains nearly
flat across the full dose range, consistent with the weak dose response
inferred from the main-text analysis.  These trends should be read as
apparent channel-level responses of fixed spectral windows, not as direct
measurements of isolated bond populations.

\section{Training Diagnostics}
\label{sec:training}

\begin{figure*}[p]
\centering
\includegraphics[width=\textwidth]{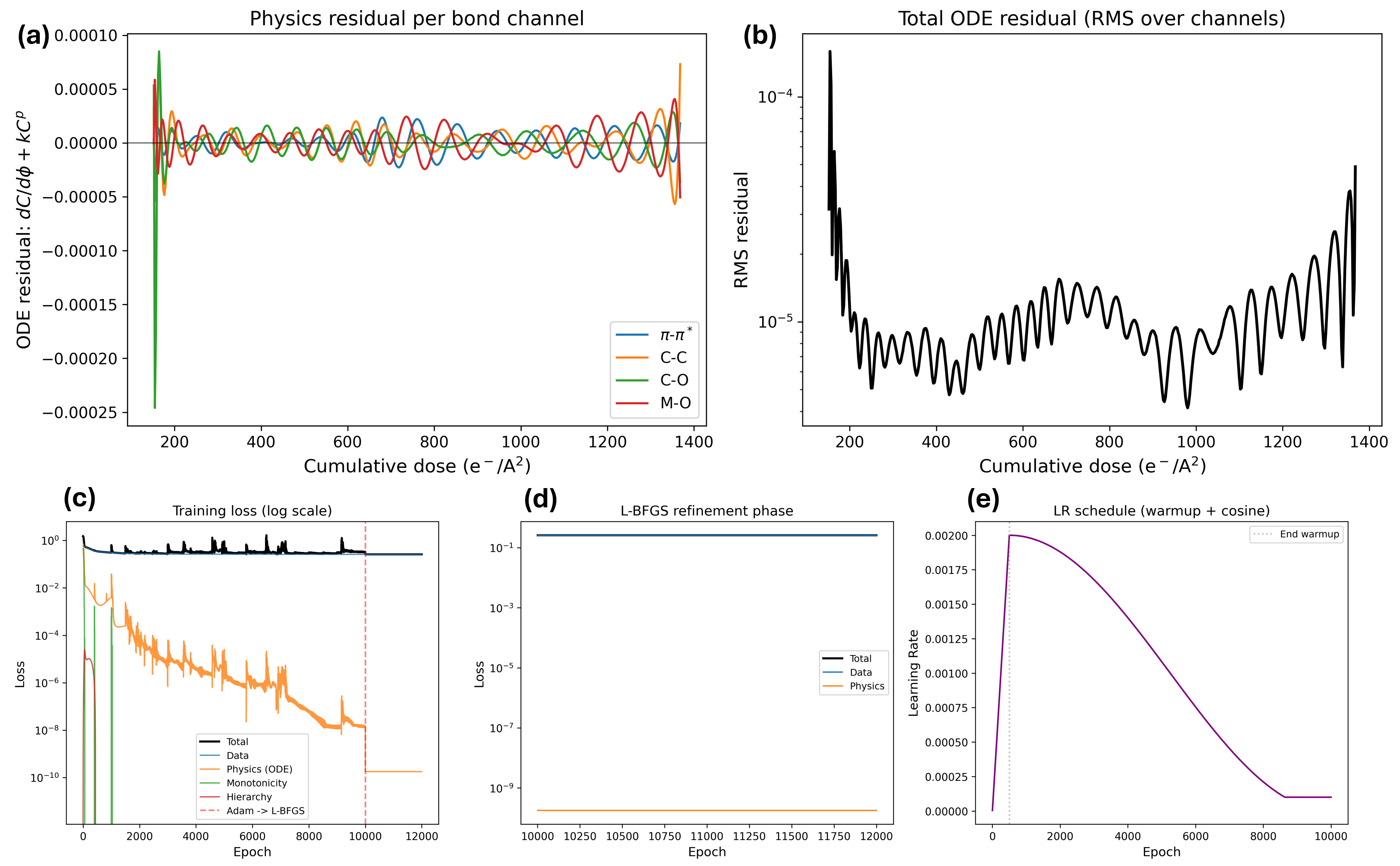}
\caption{Training diagnostics for the ensemble PINN ($N = 5$).
(a)~Physics residual ($dC_i/d\phi + k_i C_i^{p_i}$) per spectral channel
evaluated over the normalized dose domain.
(b)~Total RMS ODE residual (aggregated over all channels); the dashed
line marks the $10^{-5}$ convergence threshold.
(c)~Training loss components (log scale) during the Adam phase
(10{,}000 epochs); the dashed vertical line indicates the transition to
L-BFGS.
(d)~Data and physics loss during the L-BFGS refinement phase
(2{,}000 iterations).
(e)~Learning rate schedule: linear warmup (1{,}000 epochs) followed by
cosine annealing.}
{\small\itshape\raggedright
Alt text: Five panels showing PINN training diagnostics: per-channel physics residuals near zero, total ODE residual dropping below the convergence threshold, training loss components decreasing over Adam and L-BFGS optimization phases, and a warmup-cosine learning rate schedule.\par}
\label{fig:S2}
\end{figure*}

Figure~\ref{fig:S2} summarizes the training diagnostics for the ensemble
PINN.  The per-channel physics residuals
$dC_i/d\phi + k_i C_i^{p_i}$ [Fig.~\ref{fig:S2}(a)] remain within
$\pm 1 \times 10^{-4}$ across the normalized dose domain for all four
spectral channels after convergence, with the largest transient excursions
localized near the boundary at $\Phi_0 = 152$~e$^-$/\AA$^2$.  The total
RMS ODE residual [Fig.~\ref{fig:S2}(b)] confirms convergence below the
$10^{-5}$ threshold over the majority of the dose range, with residual
oscillations at low and high dose reflecting the finite data density at
the domain boundaries.

The training loss history [Fig.~\ref{fig:S2}(c)] shows the evolution of
data fidelity, physics (ODE), monotonicity, and hierarchy loss
components during the Adam phase (10{,}000 epochs).  The dashed vertical
line marks the transition to L-BFGS refinement, which drives the physics
and data losses down by an additional one to two orders of magnitude
[Fig.~\ref{fig:S2}(d)].  The learning rate schedule
[Fig.~\ref{fig:S2}(e)] employs a linear warmup (1{,}000 epochs)
followed by cosine annealing to a minimum learning rate of
$1 \times 10^{-5}$.

\section{Core-Loss EELS Comparison}
\label{sec:coreloss}

\begin{figure*}[p]
\centering
\includegraphics[width=\textwidth]{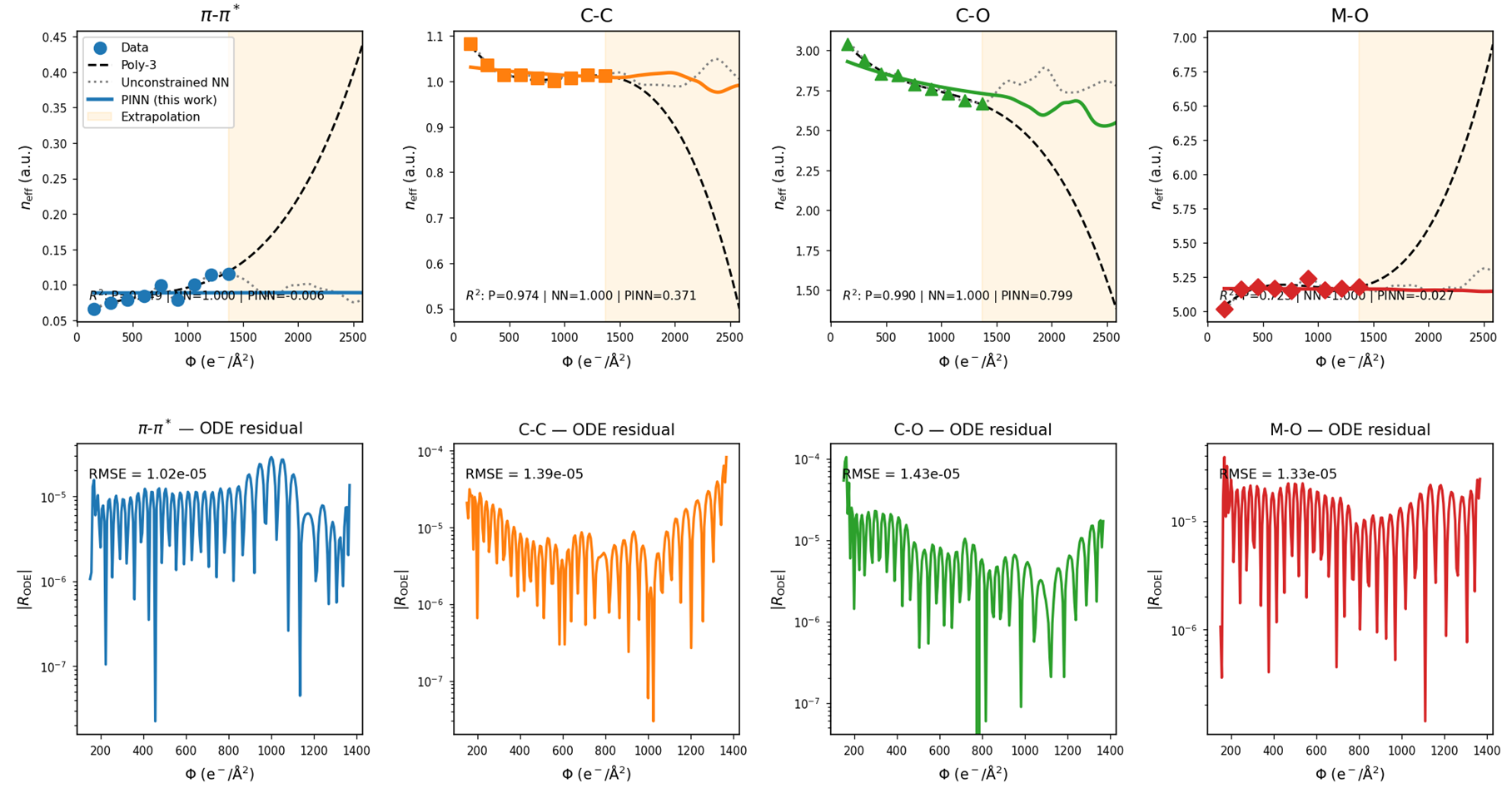}
\caption{Ablation study: PINN constrained interpretability versus unconstrained baselines ($N = 9$ dose frames). \textbf{(a)} Per-channel apparent spectral descriptors ($n_\mathrm{eff}$) versus cumulative dose $\Phi$ for a degree-3 polynomial (Poly-3), an unconstrained neural network with identical architecture but no physics loss (Unconstrained NN), and the full constrained PINN (this work). The orange region denotes the extrapolation regime. The unconstrained NN achieves $R^2 = 1.000$ across all channels by memorizing the sparse data, yet produces physically inadmissible extrapolations. The low $R^2$ of the $\pi$--$\pi^*$ ($R^2 = 0.006$) and M--O ($R^2 = 0.027$) channels reflects the dominance of the ODE constraint over data interpolation in poorly identifiable channels.
\textbf{(b)} Absolute ODE residual $|dC_i/d\phi + k_i C_i^{p_i}|$ at $N_c = 200$ collocation points. RMSE $\sim 10^{-5}$ across all channels confirms that the PINN satisfies the governing degradation ODE to high precision, demonstrating that its value lies in physics-consistent parameter extraction rather than data interpolation.}
{\small\itshape\raggedright
Alt text: Eight panels comparing PINN, polynomial, and unconstrained neural network fits to four EELS channels, plus per-channel ODE residuals. Unconstrained models fit sparse data well but extrapolate unphysically; the PINN remains bounded and satisfies the governing equation with low residual.\par}
\label{fig:S4}
\end{figure*}

To provide complementary, chemically-specific context for the fixed-window low-loss descriptors used in the main text, we acquired core-loss EELS spectrum images (C--K, O--K, and Fe--L$_{2,3}$ edges) at the lowest- and highest-dose frames of the same series (F0, 152~e$^-$/\AA$^2$; F8, 1368~e$^-$/\AA$^2$)(Fig.~\ref{fig:S4}), pooled over the full spatial region of interest and background-subtracted using an Egerton power-law fit to the pre-edge region\cite{Egerton2011}. At the C K-edge, two resolved $\pi^*$ features are labelled $\pi^*_1$ ($\sim$285~eV) and $\pi^*_2$ ($\sim$289~eV), corresponding to the aromatic-ring and carboxylate environments of the linker, respectively, following the same core-loss assignment used for terephthalate-based analogues~\cite{francis1992,Hitchcock1992}. These core-loss labels ($\pi^*_1$, $\pi^*_2$) are distinct from the fixed-window low-loss descriptors (C--C, C--O) discussed in the main text, which probe the collective dielectric response rather than bond-localized transitions. The broad feature centered near 294~eV is consistent with the overlapping $\sigma^*_{\mathrm{C-C}}$ and $\sigma^*_{\mathrm{C=O}}$ resonances reported at 293.6 and 294.7~eV, respectively, for terephthalaldehyde, a direct structural analogue of the BDC linker~\cite{Hitchcock1992}. Qualitatively, the $\pi^*_2$ (carboxylate) feature shows a loss of definition relative to $\pi^*_1$ (aromatic ring) with increasing dose, while the Fe L$_{2,3}$ white-line ratio remains visually stable, consistent with a linker-dominated, metal-node-preserving degradation pathway. Window-integrated peak areas do not show a statistically significant dose trend within the spatial-block sampling uncertainty of the present dataset ($n=8$ blocks per frame); we therefore present this comparison as qualitative, complementary evidence rather than a quantitative validation of the low-loss window assignments.

\section{Ablation Study}
\label{sec:ablation}

\begin{figure*}[p]
\centering
\includegraphics[width=\textwidth]{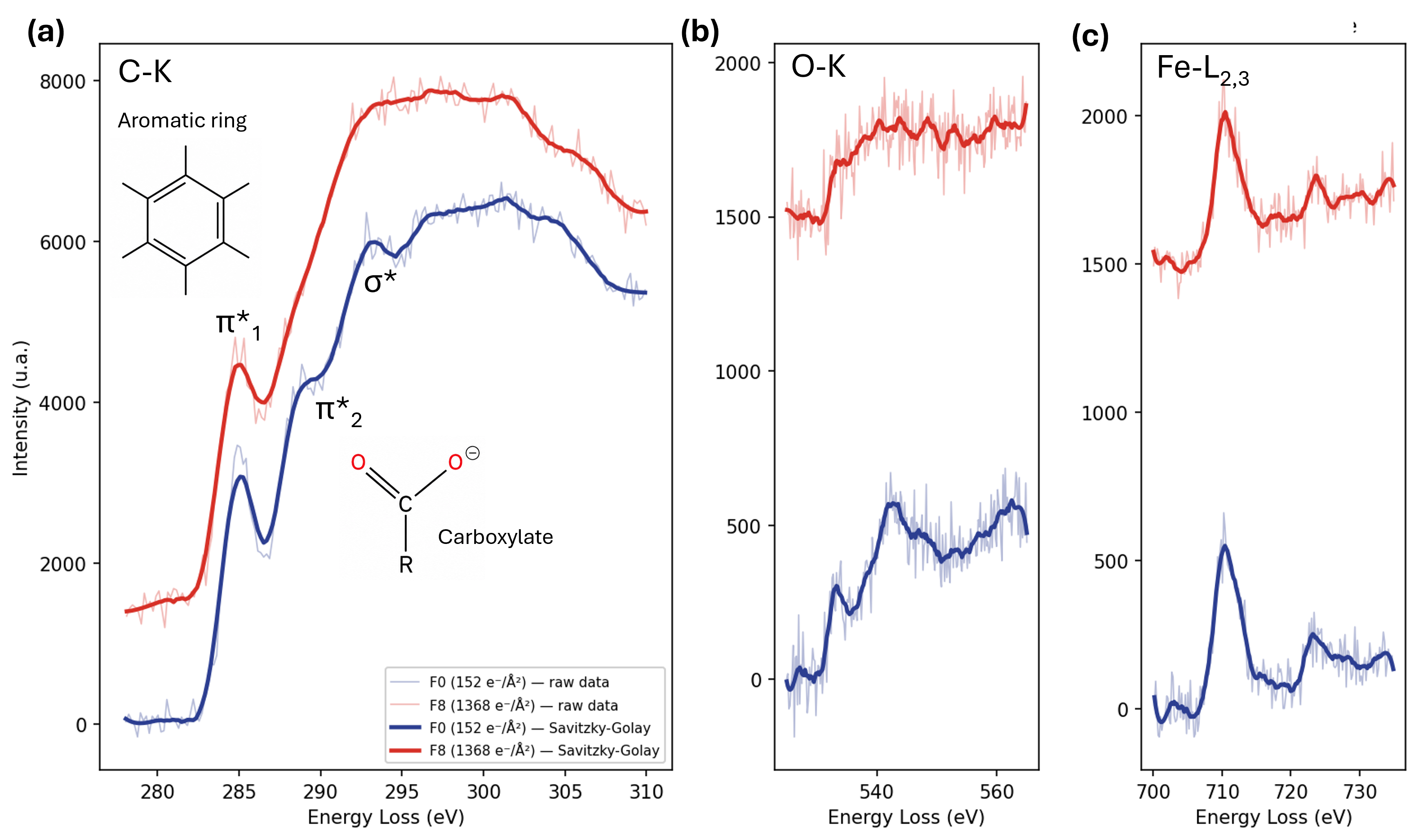}
\caption{C K-edge (a), O K-edge (b), and Fe L$_{2,3}$-edge (c) spectra for MIL-101(Fe), comparing the lowest- and highest-dose frames (F0, 152~e$^-$/\AA$^2$; F8, 1368~e$^-$/\AA$^2$). Thin lines: raw data (background-subtracted); thick lines: Savitzky--Golay-smoothed trend. F8 curves are offset vertically by 1500 (a.u.) for clarity. In (a), $\pi^*_1$ ($\sim$285~eV, aromatic ring), $\pi^*_2$ ($\sim$289~eV, carboxylate), and $\sigma^*$ ($\sim$294~eV) are labeled.}
{\small\itshape\raggedright
Alt text: Core-loss EELS spectra for carbon, oxygen, and iron edges comparing lowest and highest electron dose. The carboxylate-associated carbon peak loses definition relative to the aromatic-ring peak at high dose; the iron edge intensity ratio remains largely unchanged.\par}
\label{fig:S3}
\end{figure*}

To verify that the PINN extracts physically constrained trends rather than memorizing the $N = 9$ available dose frames, we compare the full model against two unconstrained baselines trained on identical data: a per-channel degree-3 polynomial (Poly-3) and a neural network with the same Fourier-feature/Modified-MLP architecture but with all physics penalties disabled ($\lambda_\mathrm{phys} = \lambda_\mathrm{mono} = \lambda_\mathrm{hier} = 0$, no hard boundary condition). Both baselines achieve $R^2 = 1.000$ across all channels, confirming
that interpolating $N = 9$ sparse points is trivially achievable by any sufficiently expressive model. In the extrapolation regime, however, both baselines produce physically inadmissible behavior, whereas the PINN remains bounded and monotonically non-increasing (Fig.~\ref{fig:S3}a).
The PINN satisfies the governing ODE with RMSE $\sim 10^{-5}$ at
$N_c = 200$ collocation points (Fig.~\ref{fig:S3}b) ---
a constraint inaccessible to any unconstrained model by construction. The low in-sample $R^2$ of the $\pi$--$\pi^*$ and M--O channels ($0.006$ and $0.027$, respectively) reflects the dominance of the ODE constraint over data interpolation in poorly identifiable channels, consistent with the limited-identifiability interpretation discussed in the main text.

\section{Baseline Model Comparison (AICc/BIC)}
\label{sec:comparison}
To assess whether the reported dose-dependent trends require the physics-informed
neural network or could be reproduced by simpler nonlinear fitting, we compared
three conventional baselines against each of the four low-loss window descriptors,
$\tilde{n}_{\mathrm{eff},j}(\Phi)$: a single exponential, a bi-exponential, and a
stretched-exponential (Kohlrausch--Williams--Watts, KWW) model, each fitted
independently per channel via nonlinear least squares. Given the small sample size
($n=9$ dose frames), model selection was performed using the small-sample-corrected
Akaike Information Criterion (AICc)~\cite{hurvich1989regression}, which corrects
the bias of the standard Akaike Information Criterion (AIC)~\cite{akaike1974new}
when the number of fitted parameters is a moderate-to-large fraction of the sample
size, and the Bayesian Information Criterion (BIC)~\cite{schwarz1978estimating}.
$\Delta$AICc was computed relative to the best-supported model per channel
(Table~\ref{tab:aicc_bic_baseline}). The single exponential achieves the lowest
AICc in all four channels, with $\Delta$AICc $>$ 3 relative to both alternatives
in every case, indicating that conventional bounded models are statistically
sufficient to describe the observed dose series within the present sampling.

\clearpage

\clearpage

\begin{table}[!ht]
\caption{Baseline model comparison (AICc/BIC) per channel. $\Delta$AICc computed relative to the best-supported model per channel.}
{\small\itshape\raggedright
Alt text: Table comparing exponential, stretched-exponential, and bi-exponential model fits per channel using AICc and BIC. The single exponential is best-supported for all four channels, with a clear information-criterion margin over the more complex alternatives.\par}
\label{tab:aicc_bic_baseline}
\begin{tabular}{llrrrrl}
\toprule
Channel & Model & k & AICc & $\Delta$AICc & BIC & Best \\
\midrule
$\pi$--$\pi^*$ & exponential & 3 & -78.7 & 0 & -82.9 & \checkmark \\
$\pi$--$\pi^*$ & KWW & 4 & -71.8 & 6.87 & -81 &  \\
$\pi$--$\pi^*$ & bi-exponential & 5 & -59.3 & 19.4 & -78.3 &  \\
\hline
C--C & exponential & 3 & -89 & 0 & -93.2 & \checkmark \\
C--C & KWW & 4 & -82.7 & 6.31 & -91.9 &  \\
C--C & bi-exponential & 5 & -69.8 & 19.2 & -88.8 &  \\
\hline
C--O & exponential & 3 & -66.3 & 0 & -70.6 & \checkmark \\
C--O & KWW & 4 & -63.4 & 2.99 & -72.6 &  \\
C--O & bi-exponential & 5 & -53.9 & 12.4 & -73 &  \\
\hline
M--O & exponential & 3 & -56.2 & 0 & -60.5 & \checkmark \\
M--O & KWW & 4 & -49 & 7.2 & -58.3 &  \\
M--O & bi-exponential & 5 & -37.1 & 19.2 & -56.1 &  \\
\bottomrule
\end{tabular}
\end{table}

\ifdefined\SIINCLUDED
\else
  \bibliography{references}